\documentclass{article}

\usepackage[english]{babel}

\usepackage[letterpaper,top=2cm,bottom=2cm,left=3cm,right=3cm,marginparwidth=1.75cm]{geometry}
\usepackage[utf8x]{inputenc}
\usepackage{fancyvrb}
\usepackage[T1]{fontenc}
\usepackage{booktabs} 
\usepackage{float}
\usepackage{graphicx}
\usepackage{amsmath}
\usepackage{amsfonts}
\usepackage{listings}
\usepackage{array}
\usepackage{setspace}
\usepackage{latexsym,subeqnarray,amsmath}

\usepackage{graphicx}
\usepackage{a4wide}
\usepackage{amsmath}
\usepackage{amsfonts}
\usepackage{amssymb}
\usepackage{algorithm}
\usepackage{algorithmic}
\usepackage{bbm}
\usepackage{setspace}
\usepackage{latexsym,subeqnarray,amsmath}

\usepackage{listings}
\usepackage{color}
\definecolor{gris}{rgb}{0.82,0.82,0.82}
\def\captionof#1#2{{\def\@captype{#1}#2}}

\newtheorem{example}{Example}

\floatname{algorithm}{Procedure}
\makeatletter
\def\captionof#1#2{{\def\@captype{#1}#2}}
\makeatother
\usepackage{amsmath,graphicx}
\usepackage{amsfonts}
\usepackage{amssymb}
\usepackage{graphicx}

\usepackage{amsmath}
\usepackage{graphicx}
\usepackage[colorlinks=true, allcolors=blue]{hyperref}

\title{High Performance Optimization at the Door of the Exascale}
\author{Claude Tadonki\\
 MINES ParisTech - PSL\\
 D\'epartement Math\'ematiques et Syst\`emes \\
 Centre de Recherche en Informatique (CRI)\\
35, rue Saint-Honor\'e\\
77305, Fontainebleau-Cedex\\
claude.tadonki@mines-paristech.fr
}

\begin{document}
\maketitle

\begin{abstract}
The next frontier of high performance computing is the {\em Exascale}, and this will certainly stand as a noteworthy step in the quest for processing speed potential. In fact, we always get a fraction of the technically available computing power (so-called {\em theoretical peak}), and the gap is likely to go hand-to-hand with the hardware complexity of the target system. Among the key aspects of this complexity, we have: the {\em heterogeneity} of the computing units, the {\em memory hierarchy and partitioning} including the non-uniform memory access (NUMA) configuration, and the {\em interconnect} for data exchanges among the computing nodes. Scientific investigations and cutting-edge technical activities should ideally scale-up with respect to sustained performance. The special case of quantitative approaches for solving (large-scale) problems  deserves a special focus. Indeed, most of common real-life problems, even when considering the artificial intelligence paradigm, rely on optimization techniques for the main kernels of algorithmic solutions. Mathematical programming and pure combinatorial methods are not easy to implement efficiently on large-scale supercomputers because of {\em irregular control flow}, {\em complex memory access patterns}, {\em heterogeneous kernels}, {\em numerical issues}, to name a few. We describe and examine our thoughts  from the standpoint of large-scale supercomputers.
\end{abstract}

\section{Scientific context}
The most notorious computing challenges mainly come from combinatorial problems and their applications. As the power of supercomputers is increasing, large-scale scenarios of common problems are under consideration and are expected to enter into routine. As previously stated, most of these problems can be expressed and solved in the standpoint of optimization, combinatorial and/or numerical. Serious efforts are being made to derive more powerful techniques for combinatorial,  numerical, and hybrid optimization. At this point, the word ``powerful'' refers to the complexity in terms of the number of basic steps or any relevant metric. When moving to an implementation on computers, targeting performance through efficiency turns to be a difficult task, which is exacerbated by the specific complexity and constraints of modern computing systems. The case of linear programming is particularly illustrative of what can appears as disconcerting. Indeed, the traditional {\em simplex} method, which is known to have an exponential (worst case) complexity yields more efficient implementations than the polynomial {\em ellipsoid} method. We think that similar facts will come up with large-scale optimization on {\em exascale} systems. Fundamental methods for solving problems are computer agnostic, thus,  implementation efforts mainly try to map an existing method onto a given computing system. A full and consistent optimization solution is likely to be a mix of several distinct components from the computing standpoint. Beside linear and non-linear algebra kernels, there are pure combinatorial modules, all orchestrated by at a higher level following the rules of the global method being so implemented.  
\section{Technical context}
High Performance Computing (HPC) aims at providing powerful computing solutions to scientific and real life problems. Many efforts have been made on the way to faster  supercomputers, including generic and customized configurations. The advent of multicore architectures is noticeable in the HPC history, because it has brought the underlying parallel programming concept into common considerations. Based on multicore processors, probably enhanced with acceleration units, current generation of supercomputers is rated to deliver an increasing peak performance, the {\em Exascale} era being the current horizon. However, getting a high fraction of the available peak performance is more and more difficult. The Design of an efficient code that scales well on a supercomputer is a non-trivial task. Manycore processors are now common, and the scalability issue in this context is crucial. Code optimization requires advanced programming techniques, taking into account the specificities and constraints of the target architecture. Many challenges are to be considered from the standpoints of efficiency and expected performances. The current faster supercomputer, the {\em  Supercomputer Fugaku}, has a peak of nearly 0.5 exaflops with 82\% for the sustained performance on LinPack, and the average sustained performance for the top 5 machines is 75\%. We can see that the increasing available power goes alongside with a better efficiency, most likely because of more efficient memory systems and a faster connection between the compute nodes. It is important to keep in mind that an effective HPC solution comes from a skillful combination of methods, programming, and machines \cite{hdr-tad}. The topic of {\em Optimization} is a very nice illustration of this observation because it has provided cutting-edge methods for solving (large-scale) problems, and the question of their efficient mapping onto large-scale  supercomputers is crucial and challenging. We now present an overview of the fundamental aspects of optimization, this part comes from our work\cite{hdr-tad} and is provided here in the intention of a self-contained report.

\section{Foundations and background}
Operations research is the science of decision making. The goal is to derive suitable mathematical models for practical problems and study effective methods to solve them as efficient as possible. For this purpose, {\em mathematical programming} has emerged as a strong formalism for major problems. Nowadays, due to the increasing size of the market and the pervasiveness of network services, industrial productivity and customers services should scale up with a whooping need and a higher quality requirement. In addition, the interaction between business operators has reached a noticeable level of complexity. Consequently, for well established companies, dealing with optimal decisions is critical to survive,
and the key to achieve this purpose is to exploit recent operation
research advances. The objective is to give a quick and accurate
answer to practical instances of critical decision problems. The role of operation research is also central in cutting-edge scientific investigations and technical achievements. A nice example is the application of the {\em traveling salesman problem} (TSP) on {\em logistics}, {\em genome sequencing}, {\em X-Ray crystallography}, and {\em microchips manufacturing}\cite{tsp_book}. Many other examples can be found in real-world applications\cite{ro_book}. A nice introduction of combinatorial optimization and complexity can be found in \cite{dimitriou,comb_opt}.

\begin{figure}[H]
\centering
\includegraphics[scale=0.3]{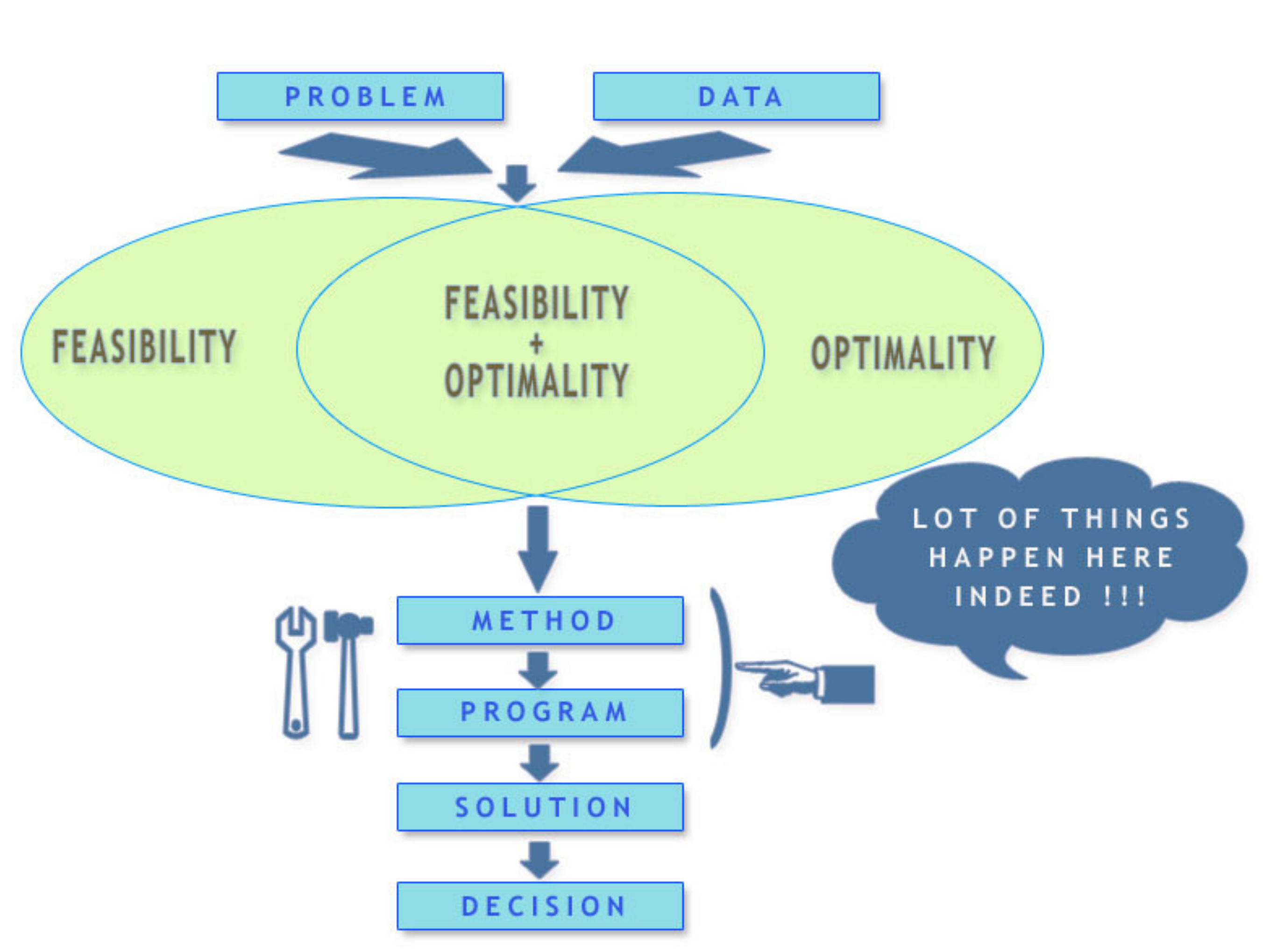}
\caption{\label{or_flow}Typical operation research workflow}
\end{figure}

The noteworthy increase of supercomputers capability has boosted the enthusiasm for solving large-scale combinatorial problems. However, we still need powerful methods to tackle those problems, and afterward provide efficient implementation on modern computing systems. We really need to seat far beyond brute force or had hoc (unless genius) approaches, as increasingly bigger instances are under genuine consideration. Figure \ref{or_flow} displays an overview of a typical workflow when it comes to solving optimization problems.   

Most of common combinatorial problems can be written in the following form
\begin{equation}
\left\{
\begin{array}{ll}
\text{minimize} & F(x)\\
\text{subject to } & P(x)\\
& x\in S,
\end{array}
\right.
\end{equation}
where $F$ is a polynomial, $P(x)$ a predicate, and $S$ the working set, generally $\{0,1\}^n$ or $\mathcal{Z}^n$. The {\em predicate} is generally referred to as {\em feasibility constraint}, while $F$ is the known as the {\em objective function}. In the case of a pure feasibility problem, $F$ could be assumed to be constant. An important class of optimization problem involves a linear objective function and linear constraints, thus the following generic formulation 
\begin{equation}
\left\{
\begin{array}{ll}
\text{minimize} & c^Tx\\
\text{subject to } & Ax\leq b\\
& x\in \mathbb{Z}^p\times \mathbb{R}^{n-p},
\end{array}
\right.
\end{equation}
where $A\in \mathbb{R}^{m\times n}$, $c\in \mathbb{R}^{n}$,   $b\in \mathbb{R}^{m}$, and $p\in \{0,1,\cdots,n\}$. If $p=0$ (resp. $p=n$), then we have a so-called {\em linear program} (resp. {\em integer linear program}), otherwise we have a {\em mixed integer program}. The corresponding acronyms are LP, ILP, and MIP respectively. In most cases, integer variables are {\em binary $0-1$ variables}. Such variables are generally used to indicate a choice. Besides linear objective functions, quadratic ones are also common, with a {\em quadratic term} proportional to $x^tQx$. We now state some illustrative examples.

\begin{example} {\bf The Knapsack Problem (KP)}\cite{kp}. The Knapsack Problem is the problem of choosing a subset of items such that the corresponding profit sum is maximized without having the weight sum to exceed a given capacity limit. For each item type $i$, either we are allowed to pick up at most $1$ (binary knapsack)\cite{kp_book}, or at most $m_i$ (bounded knapsack), or whatever quantity (unbounded knapsack). The bounded case may be formulated as follows(\ref{eqn_bkp}):\\
\begin{minipage}[b][4.2cm][t]{9cm}
\begin{equation}\label{eqn_bkp}
\left\{
\begin{array}{ll}
\text{maximize} & \displaystyle{\sum_{i=1}^{n}p_ix_i}\\
\text{subject to } & \\
& \displaystyle{\sum_{i=1}^{n}w_ix_i\leq c}\\
& x_i\leq m_i\quad i=1,2,\cdots,n\\
& x\in \{0,1\}^{n\times n}
\end{array}
\right.
\end{equation}
\end{minipage}\hfill
\includegraphics[scale=1]{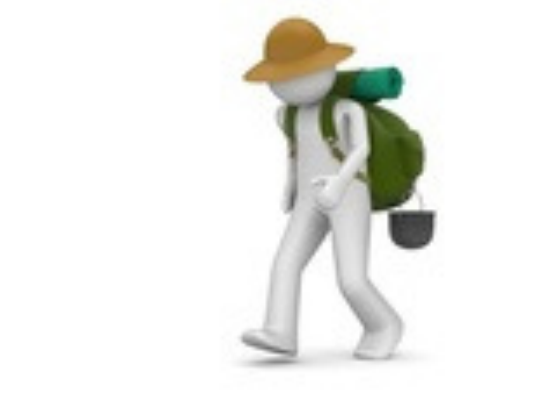}
\end{example}
\begin{example} {\bf The Traveling Salesman Problem (TSP)}\cite{tsp_book}. Given a valuated graph, the Traveling Salesman Problem is   to find a minimum cost cycle that crosses each node exactly once (tour). Without lost of generality, we can assume positive cost for every  arc and a zero cost for every disconnected pair of vertices. We formulated the problem as selecting a set of arcs (i.e. $x_{ij}\in\{0,1\}$) so as to \underline{have a tour} with a \underline{minimum cost}(\ref{eqn_tsp}). Understanding how the way constraints are formulated implies a tour is left as an exercise for the reader.\\
\begin{minipage}[b][5cm][t]{9.3cm}
\begin{equation}\label{eqn_tsp}
\left\{
\begin{array}{ll}
\text{minimize} & \displaystyle{\sum_{i=1}^{n}\sum_{j=1}^{n}c_{ij}x_{ij}}\\
\text{subject to } & \\
&\displaystyle{\sum_{j=1}^{n}x_{ij}=1}\quad i=1,\cdots,n, i\neq j\\
&\displaystyle{\sum_{i=1}^{n}x_{ij}=1}\quad j=1,\cdots,n, i\neq j \\
&x\in \{0,1\}^{n\times n}
\end{array}
\right.
\end{equation}
\end{minipage}\hfill
\includegraphics[scale=0.8]{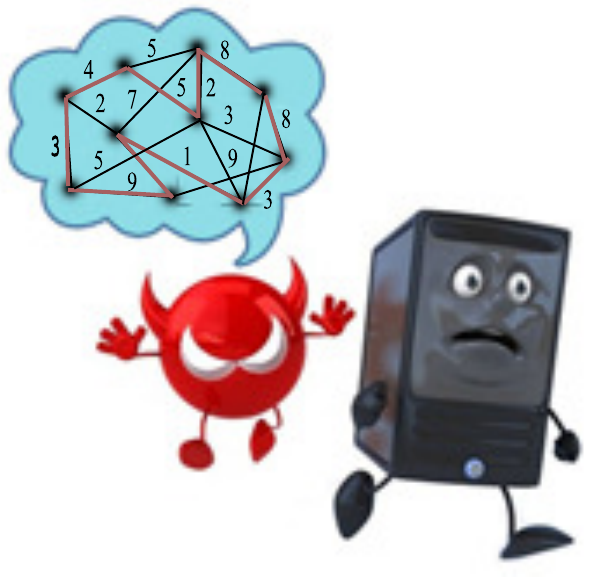}\\
The TSP has an a priori $n!$ complexity. Solving any instance with $n=25$ using the current world fastest supercomputer (FUGAKU/0.5 exaflops) might require years of calculations.
\end{example}

\begin{example} {\bf The Airline Crew Pairing Problem (ACPP)}\cite{crew}. The objective of the ACPP is to find a minimum cost assignment of flight crews to a given flight schedule. The problem can be formulated as a set partitioning problem(\ref{eqn_crew}).\\
\begin{minipage}[b][3.5cm][t]{9cm}
\begin{equation}\label{eqn_crew}
\left\{
\begin{array}{ll}
\text{minimize} & c^Tx\\
\text{subject to } & Ax=1\\
&x\in \{0,1\}^n
\end{array}
\right.
\end{equation}
 In equation (\ref{eqn_crew}), each row of $A$ represents a flight leg, while each column represents a feasible pairing. Thus, $a_{ij}$ tells whether or not flight $i$ belongs to pairing $j$.
\end{minipage}\hfill
\includegraphics[scale=0.6]{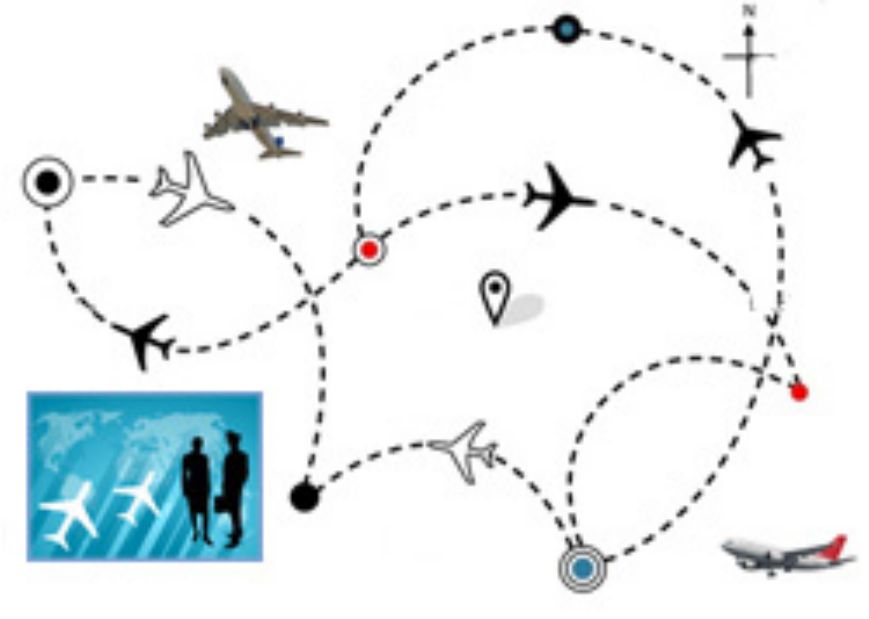} 
\end{example}

In practice, the feasibility constraint is mostly the heart of the problem. This the case for the TSP, where the feasibility itself is a difficult problem (the {\em Hamiltonian cycle}). However, there are also notorious cases where the dimension of the search space $S$ (i.e. $n$) is too large to be handled explicitly when evaluating the objective function. This is the case of the ACPP, where the number of valid pairings is too large to be included into the objective function in one time. We clearly see that we can either focus on the {\em constraints} or on {\em components} of the objective function. In both cases, the basic idea is to get rid of the part that makes the problem difficult, and then reintroduce it progressively following a given strategy. Combined with the well known {\em branch-and-bound} paradigm\cite{ref_bb}, these two approaches have led to well studied variants named {\em branch-and-cut}\cite{ref_bc} and {\em branch-and-price}\cite{ref_bp} respectively. 

\begin{figure}[H]
\centering
\includegraphics[scale=0.4]{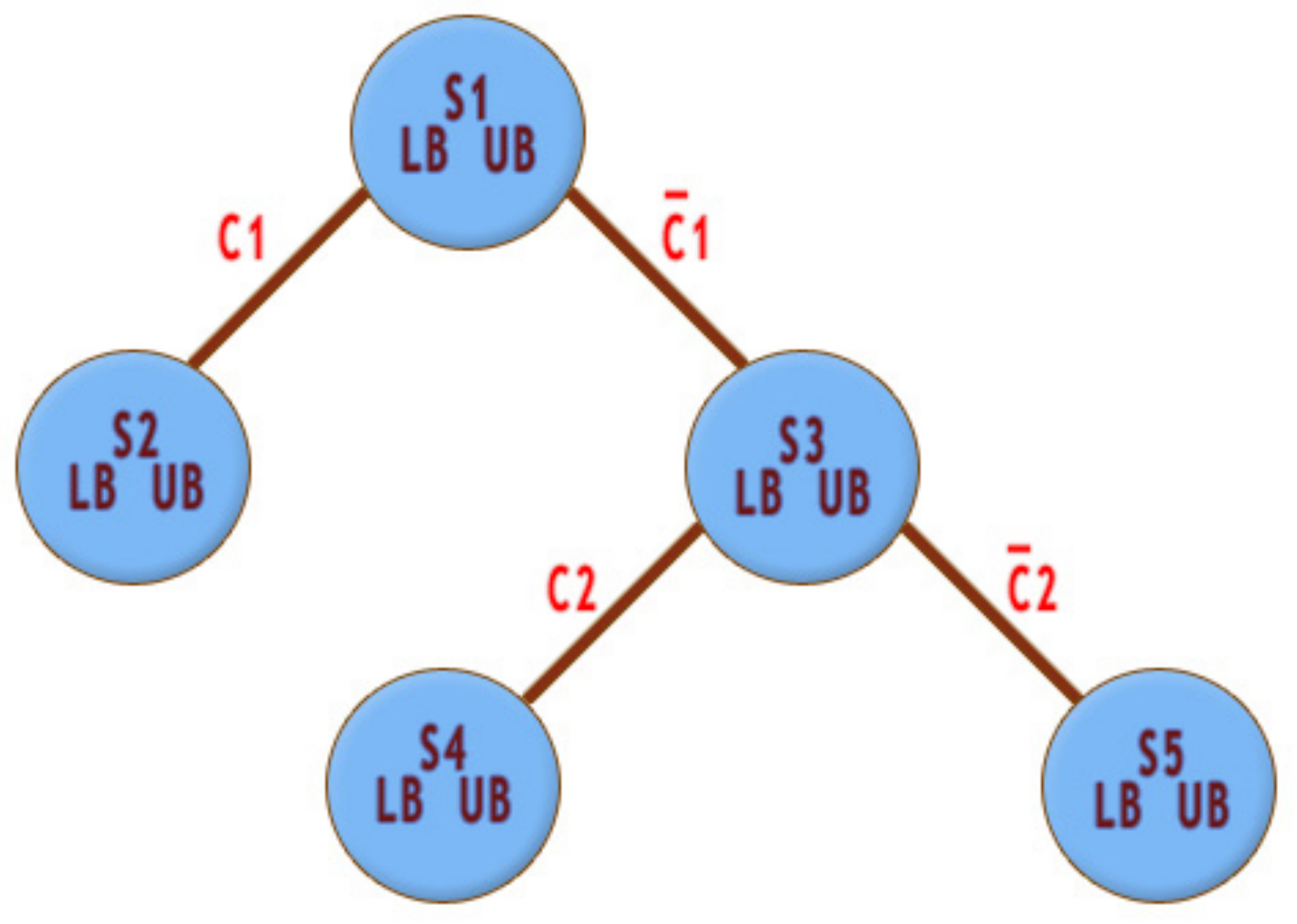}
\caption{\label{or_bb}Branch-and-bound overview}
\end{figure}

\begin{figure}
\includegraphics[scale=0.4]{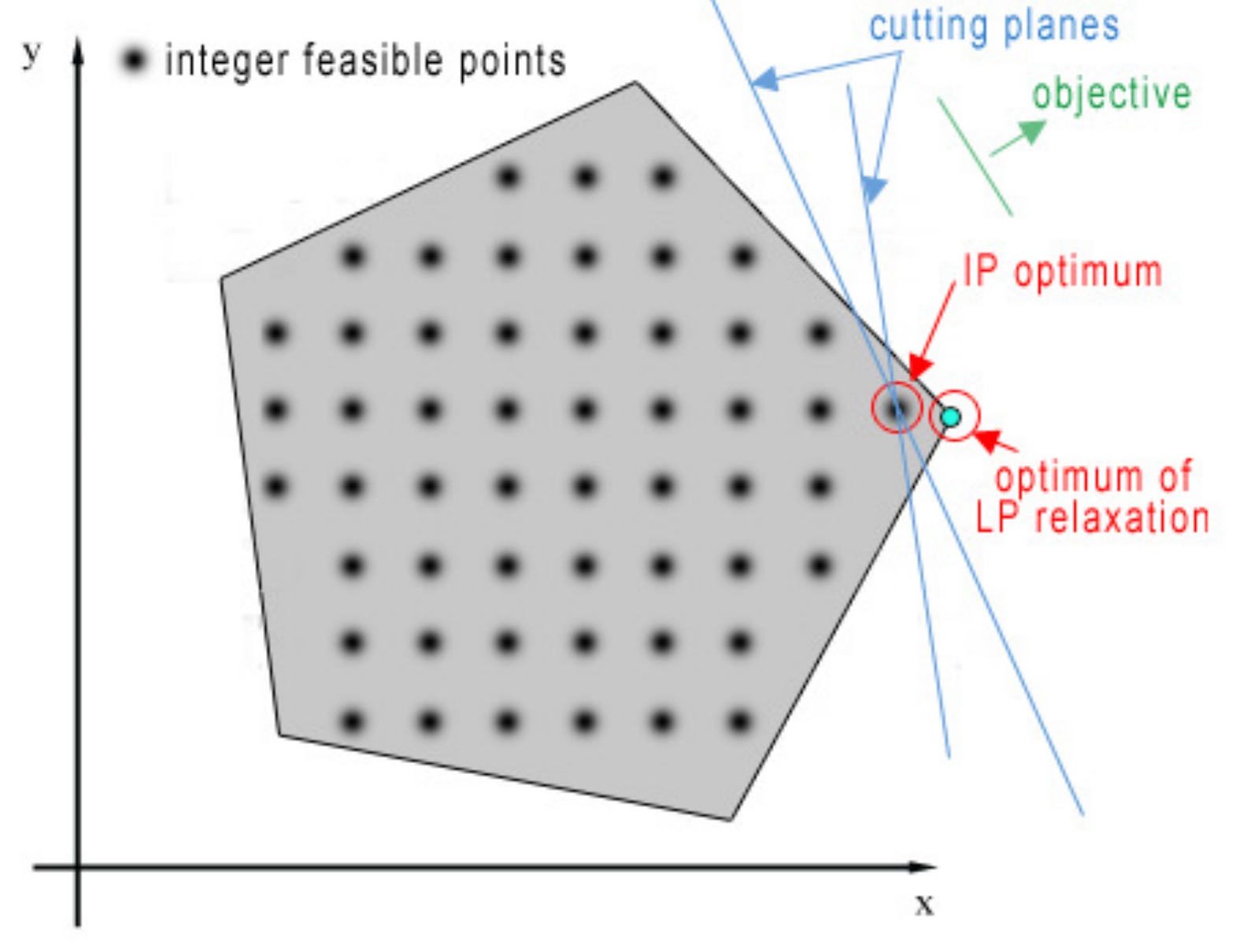}
\caption{\label{cutp_bb}Integer programming \& LP}
\end{figure}

\begin{figure}
\includegraphics[scale=0.4]{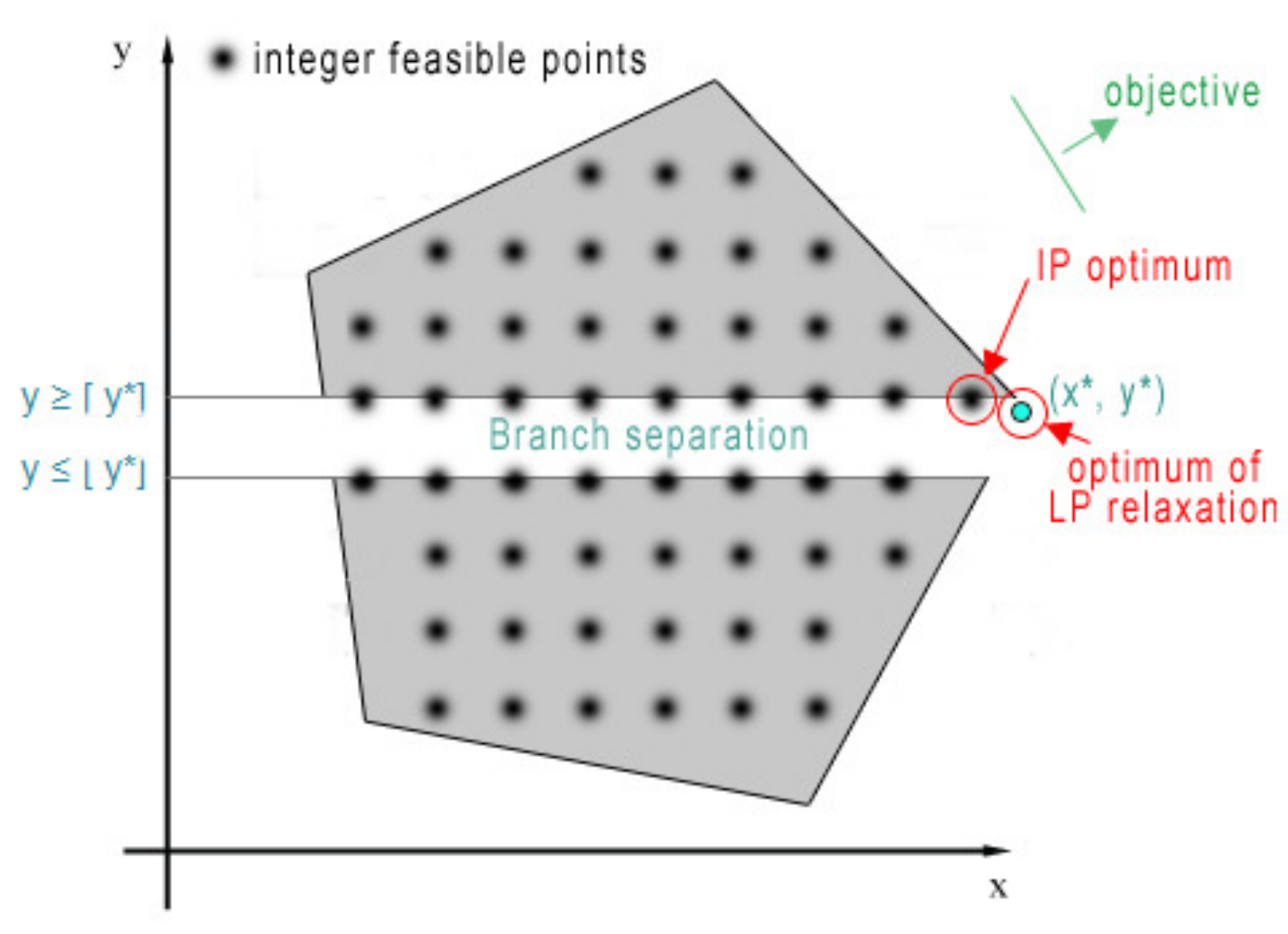}
\caption{\label{act_bb}Branch-and-bound \& LP}
\end{figure}

The key ingredient of this connection between discrete and continuous optimization is {\em linear programming} (LP). Indeed, applying a {\em linear relaxation} on the exact formulation of a combinatorial problem, which means assuming continuous variables in place of integer variables, generally leads to an LP formulation from which lower bounds can be obtained (upper bounds are obtained on feasible guests, mostly obtained through heuristics). LP is also used to handle the set of constraints in a {\em branch-and-cut}, or to guide the choice of new components ({\em columns}) of the objective function in the {\em branch-and-price} scheme. Figure \ref{cutp_bb} depicts the linear relaxation of an IP configuration, while Figure \ref{act_bb} provides a sample snapshot of an LP driven branch-and-bound.

Linear programming has been intensively studied and has reached a very mature state, even from the computing standpoint. Nowadays, very large scale LP can now be routinely solved using specialized software packages like CPLEX\cite{cplex} or MOSEK\cite{mosek}. 

\begin{figure}[H]
\centering
\includegraphics[scale=0.35]{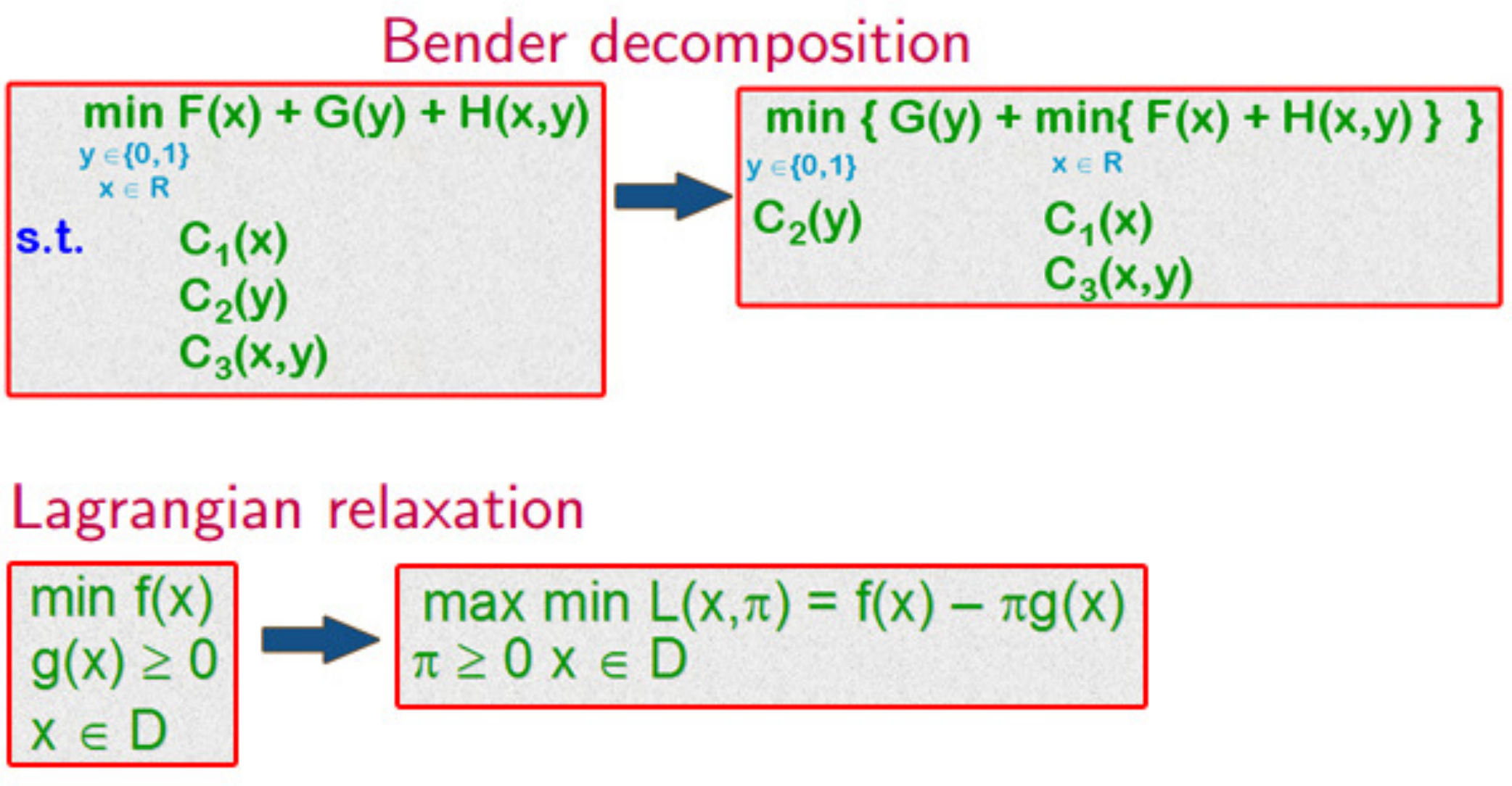}
\caption{\label{fig_bender}Bender decomposition \& Langrangian relaxation}
\end{figure}

\begin{figure}[H]
\centering
\includegraphics[scale=0.4]{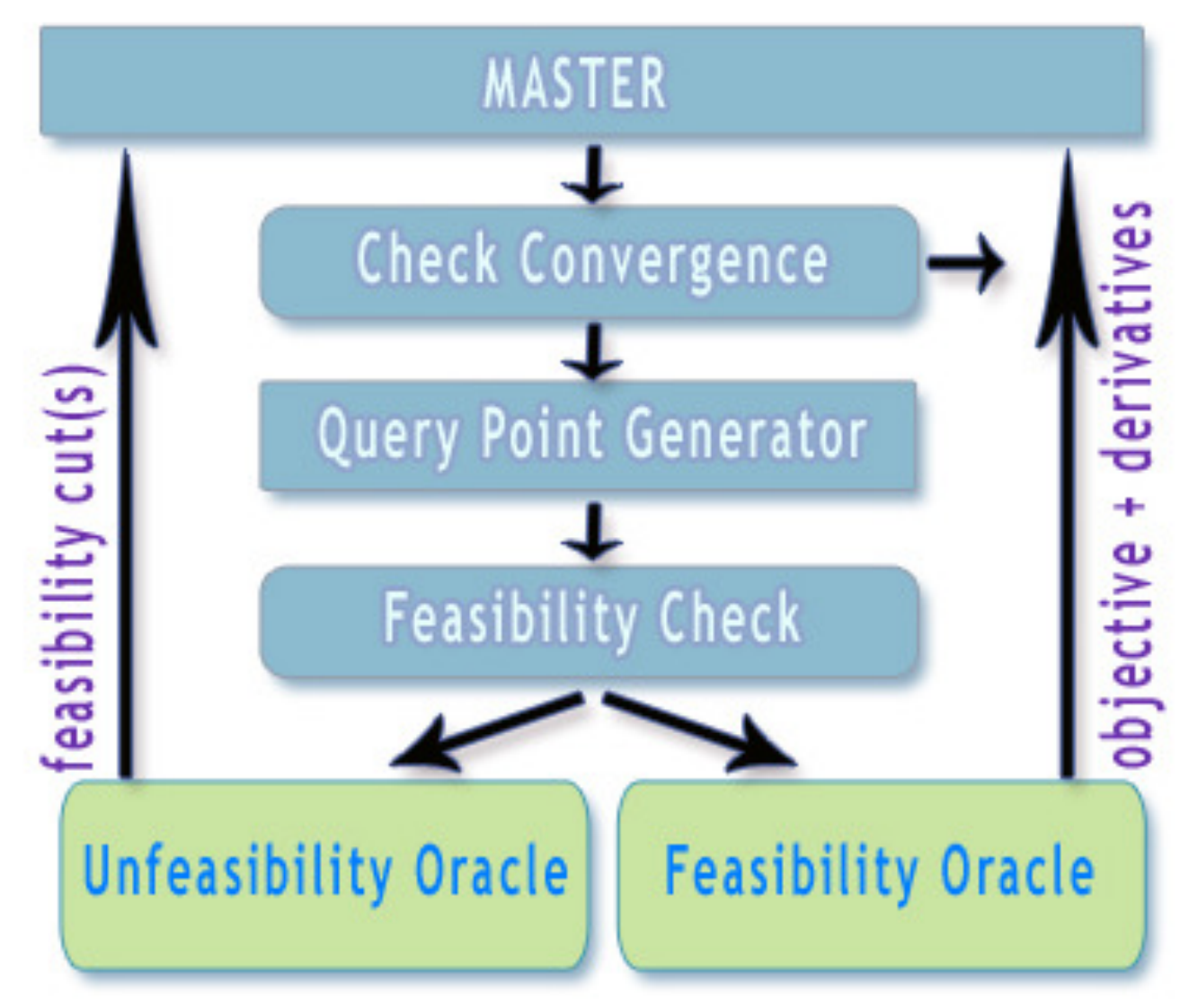}
\caption{\label{opt_oracle}Oracle based optimization workflow}
\end{figure}

{\em Branch-and-bound} and its variants can be applied to a mixed integer programming formulation by means of basic techniques like {\em Bender decomposition}\cite{bender} or {\em Lagrangian relaxation}\cite{lagrangian}. Figure \ref{fig_bender} depicts the basic idea behind these two approaches.
The later is likely to yield {\em non-differentiable optimization} (NDO) problems. Several approaches for NDO are described in the literature\cite{GoVi02}, including an oracle-based approach\cite{ref_oboe}, which we will later describe in details as it illustrates our major contribution on that topic. Figure \ref{opt_oracle} gives an overview of an oracle based mechanism.
From the methodological point of view, optimization (both continuous and combinatorial) has been so far subject to intensive and fruitful  investigations. New optimization
paradigms or improved variants of classical techniques have
reached an acceptable level of maturity, and have proved successful on number of notorious practical problems. However, in some cases,
the expected level of performance can be achieved only through parallel
implementation on large-scale supercomputers, especially with intractable (but pertinent) combinatorial
problems. The idea is to combine the advantages of mathematically
powerful algorithms with the capability of machines that have
several processors. The existence of commercial multiprocessor
computers has created substantial interest in exploring the use
of parallel processing for solving optimization problems even for basic issues. The challenge is to find a suitable way to implement the aforementioned techniques (somehow irregular) on modern supercomputers (mostly tailored for regular computations) with an acceptable efficiency. We now provide technical details on how this can be tackled and what has been done.

\section{Parallel optimization} Processing with supercomputers is mainly  parallel computing. Theoretical complexity studies the intrinsic difficulty of the optimization problems and classify them accordingly. There is an important set of common problems that can be solved or approximated in polynomial time. However, as some of them are (recursively) solved to get the solution of more difficult problems, improvements are still expected whenever there is a room for that. A good example of this is the {\em shortest paths problem}, which appears as a sub-problem for the {\em multicommodity flow problem}\cite{mcf}. Many other combinatorial problems are (known to be) difficult, thus the basic expectation with supercomputers is to be able to solve them in a reasonable time through an efficient parallelization of a chosen method. The main point with large-scale supercomputers (thus exascale ones) is the huge number of computing units, which implies a larger and deeper parallelism. 
The global topic of Optimization has mainly two components: {\em continuous optimization} and {\em discrete
optimization}. However, because pure combinatorial problems might be too difficult to solve only from the combinatorial standpoint, many approaches have developed a bridge between the discrete universe and
the continuous universe through geometric, analytic, and algebraic
techniques such as {\em global optimization}, {\em semidefinite
programming}, and {\em spectral theory}. Mixed integer
programming formulations  involve differentiable or
non-differentiable objective functions. Non-differentiable configurations might come from the consideration of a {\em Lagrangian relaxation} approach, which moves a subset of the constraints (usually the harder ones) into the objective function. 
The efforts in the design of efficient algorithms for common combinatorial problems has lead to useful connections among
problems (i.e. a solution for one can be used to construct a solution for another). As consequence, there is a set of
reference optimization problems for which improved solutions are
continuously tracked by researchers. For this purpose, {\em
parallel computing} applied to all previously mentioned optimization paradigms is clearly worth considering.

An important set of discrete optimization problems are NP-
complete\cite{GAREY}; hence their time complexity increases
exponentially for all known algorithms. Consequently, parallel
processing cannot achieve polynomial complexity on these problems without using at
least an exponential number of processors (not counting data exchanges). However, the
average-time complexity of heuristics and sub-optimal algorithms
for a wide range of problems are polynomial\cite{Pearl1, Wah1}.
Significant advances have been made in the use of powerful
heuristics and parallel processing to solve large scale discrete
optimization problems. Number of problem instances that were considered
computationally intractable on sequential machines are 
routinely solved on server-class symmetric multi-processors and workstation clusters. In conjunction with the increasing power of supercomputers, cutting-edge methods in optimization are expected to cope with very large-scale problems.  

We get a direct impact of parallel computing in numerical optimization through the advances in parallel numerical algebra\cite{alpa1,alpa2,alpa3}, with some of them being implemented into effective frameworks\cite{par_tool_1,par_tool_2,par_tool_3,par_tool_4,par_tool_5}. Encapsulating parallel linear algebra routines into optimization codes \cite{Ap1, pcplex} is a nice way to provide their power to the users without additional efforts. This is still a very critical and challenging topic since parallelizing the linear algebra kernels of optimization algorithms is not an easy task, and moving on the exascale era will made it more complex. For instance, matrix factorization updating process in quasi-Newton methods or active set strategies involves vector-vector operations that are not easy to parallelize efficiently \cite{Ap2}. According to Schnabel\cite{Scha1}, parallel computing can be performed in numerical optimization through three levels:  
\begin{itemize}
\item
parallelization of the function and/or the derivative evaluations;
\item
parallelization of the linear algebra kernels;
\item
modifications of the basic algorithms in order to increase the degree of parallelism.
\end{itemize}

For the first two levels, several approaches have been developed \cite{alpa1,pard1, pard2,pard3,gonz1} and we might also expect some outputs for heterogeneous systems. For most of interior point (IP) methods in linear programming (LP), quadratic programming (QP), and nonlinear programming, the kernel is the solution of a special linear system \cite{and1,ben1}. As the iterates approach the boundary of the feasible set or the optimal solution, the system becomes more and more ill-conditioned. Suitable strategies have been developed within a modified Cholesky factorization framework and successfully used in specialized codes as CPLEX\cite{cplex},  LOQO\cite{loqo} and GUROBI\cite{gurobi}. Thus, efficient parallel versions of these strategies are highly desired, but challenging on large-scale supercomputers, especially for sparse systems. The paper of Durazzi and Ruggiero \cite{Dur1} presents a parallel approximated IP method for QP, based on a preconditioned Conjugated Gradient algorithm. D'Apuzzo and Marino \cite{Ap2} have proposed a parallel Potential Reduction algorithm for the convex quadratic bound constrained problem. A parallel decomposition approach is considered by Zanghirati and Zanni \cite{za1} for large scale QPs. Blomwall \cite{blo1} has proposed a parallel implementation of a Riccati-based primal IP algorithm for multistage stochastic programming. Most of these contributions together with more recent ones consider conventional parallel architectures, the question is how well are they adaptable for complex systems (i.e. heterogeneous with NUMA memory for instance).

Regarding the third level, multi-directional search strategies \cite{lew1} provide a high-level parallelism which can be exploited through a concurrent execution of the minimization processes. Ad-hoc or application specific algorithms are also concerned, particularly when large-scale instances are considered \cite{By1, Joo1}. Another case study in statistical model selection is analyzed by Gatu and Kontoghiorghes \cite{Ga1}. As many fields in numerical analysis, several algorithms in numerical optimization have been revisited because of parallelism considerations. In \cite{part_ferris},  approaches to expose parallelism through appropriate partitioning of mathematical programs are reported. Interior point strategies, because  of their direct possibility of parallel implementation \cite{az2, gup1}, have received much attention  compare to active set algorithms, and have stimulated intensive researches in order to understand and overcome their weak scaling on large supercomputers. Developments in object oriented software for coding and tuning linear algebra algorithms at a high level of abstraction are provided in \cite{va1, wha1}

As previously said, many techniques have so far been developed to provide a bridge between continuous and discrete formulations. Recent successes based on such approaches include IP methods for discrete problems, the Goemans-Williamson relaxation of the maximum cut problem, the Chvatal cuts for the traveling salesman problem, and the Gilbert-Pollak's conjecture, to name a few. Parallel algorithms for discrete optimization problems can be obtained in many different ways including the classical {\em domain decomposition}.  SPMD (Single Program Multiple Data) parallelization attempts to enlarge the exploration of the solution space by initiating multiple simultaneous searches towards the optimal solution. These approaches are well implemented by clustering methods. Byrd et al. \cite{By1, By2} and Smith and Schnabel\cite{smt1} have developed several parallel implementations of the clustering method. Parallelization of classical paradigms have also been explored: {\em parallel dynamic programming}\cite{ge1}, {\em branch and bound}\cite{ref_bb, abb}, {\em tabu search, simulated annealing}, and {\em genetic algorithms}\cite{la1}. In the paper of Clementi, Rolim, and Urland \cite{cle1}, randomized parallel algorithms are studied for {\em shortest paths, maximum flows, maximum independent set}, and {\em matching problems}. A survey of parallel search techniques for discrete optimization problems are presented in \cite{Gram1}. The most active topics are those involved with searching over trees, mainly the {\em depth-first} and the {\em best-first} techniques and their variants. The use of parallel search algorithms in games implementation has been particularly successful, the case of IBM's Deep Blue \cite{ham1} is illustrative. This topic is very active in {\em Artificial Intelligence} for which the question of efficient parallelization stands as one of the major HPC applications. 

We now discuss what appears to us crucial points on the way to the Exascale when considering efficient implementations in both continuous optimization and combinatorial optimization.
\clearpage
\section{Critical Numerical and Performance Challenges}
Let first point out and describe the main issues when seeking an efficient implementation of the aforementioned paradigms on large-scale supercomputers.
\begin{itemize}
 \item {\em Computing unit:} The generic compute node is likely to be a many-core processor. Seeking efficiency and scalability with many-core processors is a hard task \cite{manycore-tad}. As with any shared-memory system, the way to go is through the shared-memory paradigm. Thereby, we avoid explicit data exchanges, but there is more pressure on main memory accesses with a heavy concurrency that will be the main culprit of weak scalability. Vectorization is to be considered at the level of the linear algebra kernels, and this requires a suitable data organization.
  \item {\em Memory system:} One critical point here is the management of shared variables. Optimization techniques are likely to be iterative, so the access to these variables is repeated accordingly. For read-only accesses, the performance will depend on how good we are with memory caching, this aspect should be investigated deeply considering all iteration levels. For write accesses, the main issue is concurrency, with a special focus on iterative (in-place) updates. The case of non uniform memory access (NUMA) needs a special attention as most of many-core processors follow this specificity along with the corresponding packaging of the cores. It is likely that exascale machines will be equipped with such processors.
  \item {\em Numerical sensitivity:} A part from accuracy concerns, it is common to consider a lower precision in order to reach a faster flops through wider SIMD and also to speed-up the memory accesses. Iterative methods usually consider this adaptation under a global mixed-precision scheme. The main drawbacks with lower precision come from the potential lost of accuracy, which might led to wrong numerical results or slower convergence.
 \item {\em Heterogeneity:} The tendency with top class supercomputers is heterogeneity. The most common configuration is the classical CPU-GPU conjunction. GPUs has reached enough maturity to be considered for most of common computing tasks including those form linear algebra. It is likely that this will be and remain the typical scenario of the GPU consideration for the implementation of optimization algorithms. However, the well-known problem of CPU/GPU data exchanges is to be seriously considered in the standpoint of an iterative process. In case of high-precision computation with GPU, there might be some concerns about accuracy.  In addition, a trade-off between accuracy and performance should be skillfully considered. 
 \item {\em Synchronization:} Considering the notorious case of the {\em brand-and-bound}, which also stands as a typical connection between continuous optimization and combinatorial optimization, all active explorations running in parallel share some common variables (concurrent updates, critical values, ...) and conditions (termination/pruning, numerical/structural, ...). Synchronize in the context of a large-scale supercomputer is costly and the effect on the scalability is noteworthy.     
 \item {\em Data exchanges:} This is the main source of a serious time overhead with distributed memory parallelism, which also generally includes the aforementioned mechanism (synchronization). A more general optimization scheme has several levels of iteration (of different natures), which yields a complex communication flow and topology. This aspect is certainly the most hindering on the way to parallel efficiency, as it consumes the major part of the global overhead.  
  \item {\em Load balance:} Active subproblems might have different complexities and numerical characteristics, thus yielding unequal loads for the corresponding tasks. Beside the computing load, there is also some numerical characteristics that might impact the local runtime complexity on the compute nodes.  This aspect is hard to fix without changing the way the  computation is organized. The way a given (sub)problem is solved in optimization sometimes depends on its specific structure, this makes difficult to predict the choice that will be made at runtime and thus complicates any prediction.
\end{itemize}

Many optimization problems are based on an objective function that is implicit or non-differentiable. To solve them with gradients-based approaches, we need to deal with an {\em oracle} that can return for a given point of the search space the evaluation of the objective function and the corresponding derivatives. Oracle-based Optimization is a powerful tool for general purpose optimization. To make this approach successful from the performance and numerical standpoints, it is important to  (i) keep the number of calls to the oracle as low as possible, especially if it involves solving a difficult combinatorial problem; and (ii) take care of numerical issues that might extend the number of necessary iterations or lead to divergence. Oracle-based continuous optimization is better addressed with generic approaches so as to offer the possibility to treat most common combinatorial problems. However, number of important aspects still need to be seriously considered. We list some of them.
\begin{itemize}
 \item The core of an oracle-based method in continuous optimization involves solving a linear system for the search direction used to get the next query point. It is crucial to have the solution accurate enough to be meaningful and keep us to the track. As we get close to the boundaries of the search space or to the optimal solution, the principal matrix of the linear system becomes ill-conditioned, thus making difficult the computation of the required solution. This fact severely increases the associated computational cost, unless we chose to sacrifice the accuracy, which will extend the number of outer iterations towards the solution. Thus, it is important to carefully address this issue, which belongs to the more general topic of solving ill-conditioned linear systems. However, there is probably a way to exploit the specific structure of the principal matrix in this case. The topic here is mainly that of {\em linear systems solving}, which has been extensively studied but remains difficult to make it as scalable as desired, especially with sophisticated iterative methods. Inter-processors communication and global synchronization mechanisms are what we should care about for this aspect on {\em exascale} systems. 
 \item
The query point generator of the cutting planes method looks for a guess within the localization set that corresponds to the polyhedral defined by the cuts accumulated so far. A good management of these cuts is crucial and their number linearly increases with the number of iterations. If the dimension of the problem is huge, or if we have already performed a large number of iterations, then the required memory space necessary to keep all the cuts will become significant, and this might slowdown the global memory efficiency, especially with complex memory systems like NUMA ones. One way to fight against this problem is to eliminate redundant cuts, or to keep only the minimal set of the cuts that corresponds to the same (or equivalent) polyhedral of the localization set. Doing this is not trivial as there are many valid selections. Another way is to aggregate the cuts instead of eliminating them. We could also weight the cuts according to their importance within the localization set. All theses thoughts have to be studied deeply, even through an experimental approach. However, we need to be careful as we could destroy the coherence of the localization set, thus impacting the convergence.   
 \item
Cutting planes methods are iterative, and the convergence is monitored by the calculation of the gap between the best solution found so far and the estimated lower/upper bound (ideally the optimal value of the objective function, but we don't have it). The process converges if: (a) the gap is below the tolerance threshold; (b) we have reached the maximum number of iterations (over the expectation); (c) a null gradient is returned by the oracle;  (d) an incoherent information is provided by the oracle or calculated internally; (e) an unexpected critical issue (hardware/system or numerical) has occurred. The main focus here is the lower/upper bound estimation. This is usually obtained from the localization set (the cuts + the objectives), which might become heavy and numerically sensitive over the time (again, because of the large number of collected cuts and the global configuration). If the estimation of the lower/upper bound is good enough, then we will perform more additional iterations or never converge (even if we should, either because we are already at the optimum or there is no further improvement). It is therefore important to address this problem and look for a robust approach. It makes sense to assign this calculation to single computing unit and broadcast the result to the whole computing system.
\item
Regarding the Newton linear system that is solved during inner iterations to get the search direction for the next query point, updating the principal matrix takes a serious overhead. Indeed, at each iteration, the matrix of the  generated cuts is updated from $A$ to $[A,u]$, where $u$ is the new incoming cut, then we solve a linear system based on a principal matrix of the form
\begin{equation}\label{nmatrix}
 A \times diag(s^2)\times A^T,
\end{equation}
 where $s$ is the vector of the so-called {\em slack variables} and $s^2 = (s_i^2)$. It is quite frustrating to solve this system from scratch at each iteration. Indeed,  the principal matrix (\ref{nmatrix}) seems to have a suitable form for a direct Cholesky factorization. The dream here is to keep on the desired factorization by means of efficient updates, thus a quadratic complexity instead of the cubic one for the  factorization restarted from scratch. The current state-of-the-art in matrix computation, to the best of our knowledge, does not provides the aforementioned Cholesky update, this remains to be investigated including the parallelization from the perspective of running on a (large-scale) supercomputer. 
 \item
About the branch-and-bound, an important method for solving combinatorial problems (including approximations), very popular for MIP formulations, the main research direction from the HPC standpoint is through an efficient parallelization of the paradigm itself. Branch-and-bound is likely to yield an irregular computation scheme with an unpredictable path to the solution, thus making very challenging for efficient parallelization, especially on large-scale supercomputers. Among critical issues, we mention: {\em heavy synchronization, irregular communication pattern, huge amount of memory to handle the generated cuts}, {\em load unbalanced} and/or {\em non-regular memory accesses}. The management of the implicit recursion of the whole is difficult to keep scalable as the number of processors increases. An on-the-fly rescheduling of the tasks might be necessary at some points in order to adapt to branching mispredictions or severe load imbalance. In conjunction with continuous optimization, there are effective  generic frameworks for the branch-and-bound associated with continuous optimization solvers \cite{Briant01}, such frameworks should be made parallel at design time. 
\end{itemize}
 \section{Conclusion}
 Optimisation is a central topic, which combines advances in applied mathematics and technical computing. Powerful methods have been developed and are still improved to solve difficult but relevant real-life problems. As the power of supercomputers is significantly increasing, there is an instinctive desire for being able to routinely solve large-scale problems. This raises the challenge of efficient implementations of cutting-edge optimization techniques on large-scale supercomputers. Parallel optimization is the main topic involved in this context, and the main concern is scalability. Ideally, the most powerful optimization methods should be scalable enough to yield the most efficient solutions to the target problems. However, the global and internal structures of modern supercomputers make them not easy to program efficiently, especially with too specific approaches like the ones from optimization. On the way to the exascale, this will be exacerbated by the complexity of the systems, but the efforts are worth it. 
 
%
%
\bibliographystyle{latex8}

\end{document}